# Deep Recurrent Neural Network for Protein Function Prediction from Sequence

Xueliang Leon Liu[1,2,3]

[1] Wyss Institute for Biologically Inspired Engineering;
[2] School of Engineering and Applied Sciences, Harvard University;
[3] Department of Systems Biology, Harvard Medical School;

## Email Addresses

Xueliang Leon Liu: xliu@fas.harvard.edu

## Corresponding Author

Xueliang Leon Liu
Harvard University
Phone: 626-215-5288
Email: xliu@fas.harvard.edu

**Abstract**

As high-throughput biological sequencing becomes faster and cheaper, the need to extract useful information from sequencing becomes ever more paramount, often limited by low-throughput experimental characterizations. For proteins, accurate prediction of their functions directly from their primary amino-acid sequences has been a long standing challenge. Here, machine learning using artificial recurrent neural networks (RNN) was applied towards classification of protein function directly from primary sequence without sequence alignment, heuristic scoring or feature engineering. The RNN models containing long-short-term-memory (LSTM) units trained on public, annotated datasets from UniProt achieved high performance for in-class prediction of four important protein functions tested, particularly compared to other machine learning algorithms using sequence-derived protein features. RNN models were used also for out-of-class predictions of phylogenetically distinct protein families with similar functions, including proteins of the CRISPR-associated nuclease, ferritin-like iron storage and cytochrome P450 families. Applying the trained RNN models on the partially unannotated UniRef100 database predicted not only candidates validated by existing annotations but also currently unannotated sequences. Some RNN predictions for the ferritin-like iron sequestering function were experimentally validated, even though their sequences differ significantly from known, characterized proteins and from each other and cannot be easily predicted using popular bioinformatics methods. As sequencing and experimental characterization data increases rapidly, the machine-learning approach based on RNN could be useful for discovery and prediction of homologues for a wide range of protein functions.

**Introduction**

As the cost of DNA sequencing is decreasing drastically over the last decade, the volume of biological sequences particularly for new proteins is also increasing rapidly. Discovering the functions of these new proteins not only could allow one to better understand their roles in their native contexts, but also utilize them in synthetic biology to assembled new biological circuits and pathways for useful applications such as production of valuable compounds or treating disease. However, the experimental characterization of proteins' properties such as structure and function can be slow and resource-demanding using techniques such as x-ray crystallography, cryo-TEM, or functional assays, significantly outpaced by sequencing. A predictive pipeline that can accurately translate primary sequence to function would allow filtering of the vast sequence dataset to an experimentally manageable subset of high-confidence candidates of highest interest toward a particular function or application is greatly desired.

Currently there are several popular methods for extracting useful information from primary sequences and infer functional information based on comparison of new sequences to existing sequences of known function. For example, BLAST performs sequence alignment with heuristic scoring. Multiple sequence alignments can be used to build models that capture conservation patterns (i.e. profiles or motifs), such as Position Specific Score Matrices (PSSM)[1] or Hidden Markov Models[2]. These profiles can be used to iteratively search in a database search (e.g. PSI-BLAST, jackhmmer) to detect remote homologies, allowing the discovery of protein clusters or families that are evolutionarily related. New query sequences may be aligned to existing profiles for identification of function. The alignment scores and "E-value" can help indicate the degree of homology between the new sequence and existing sequences. Powerful and popular as these existing approaches are for protein function annotation directly from sequence, there may still be limited in classifying sequences coding for proteins with similar function or structure but are very distant in evolutionary scale or have come to adopt similar function via convergent evolution. For instance the proteases have independently evolved the "catalytic triad" active site in 23 protein superfamilies. The "catalytic triad" consists an acid-base-nucleophile configuration of three amino acids arranged in spatial proximity but can be distant on the sequence. Given the difficulty in accurate prediction of three-dimensional protein folding, the "catalytic triad" is difficult to predict based on sequence alignment. Here I present the application of machine learning using recurrent neural networks, recently gaining popularity and successes for natural languages processing, to capture high-dimensional, complex patterns in biological sequences in order to predict protein functions, potentially beyond the capability of current methods.

**Results**

*Model Architecture*

The deep learning recurrent neural network (RNN) model for protein function prediction is trained on a large set of protein sequences with certain known functions as labels. The training process tunes the parameters of the network by minimizing prediction errors (categorical entropy). After validating good prediction performance of the trained network using a "test" dataset of randomly chosen sequences of proteins with known functions but have never been seen during training, new sequences with unknown functions are fed to the network to make predictions of function (Figure 1a). The predicted function is eventually validated by experimental assay. Furthermore, RNN models could predict certain phylogenetically distinct "out-of-class" protein families with similar function, albeit with worse sensitivity and selectivity.

The recurrent neural network (RNN) model contains one or more sets of "bi-directional" recurrent layers with long-short term memory (LSTM) neurons processing the input sequence one residue or character at a time (Figure 1b). The forward layer scans the protein sequences from the N- towards the C-terminus and reversed for the backward layer, allowing the network to make use of context on both sides of each position rather than just what was seen before in a single direction. Each residue in the input protein sequence is converted into a "one-hot" vector whose elements are all 0 except at the position of the amino acid it corresponds to, where it is set to 1. Each LSTM neuron in each recurrent layer uses input "i", output "o", gate "g", and forget "f" gates to modulate the input vector and update the neuron's internal cell state "c" and hidden state "h". The gates apply matrices, whose elements are adjustable parameters to be learned, on the input and hidden state vectors at each recurrent step/layer and subsequently normalize the results with nonlinear activation functions (Methods). Intuitively, given each new input vector (i.e. sequence residue), the gates control what and how much to add to and output from the hidden state memory, which encodes sequence patterns relevant toward particular protein function (Figure 1c). These features of the LSTM architecture allow the RNN to maintain, over many recurrent iterations, the magnitudes of both the relevant signals in feed-forward propagation as well as the error gradients in backpropagation, thereby resolving the issues of loss of contextual memory and vanishing/exploding gradients that have limited the usefulness of traditional RNNs in processing long sequences (e.g. hundreds of units/iterations). The outputs from the last LSTM neurons of the forward and reverse hidden layers are eventually fed to a fully-connected layer of artificial neurons, where each neuron represents one functional class and outputs via the "softmax" activation function the probability that the input sequence represents a particular functional class. The number of recurrent hidden layers, LSTM neurons in each layer, hidden units (i.e. hidden state vector dimension) in each LSTM neuron, and the architecture of each LSTM neuron (e.g. "peep-hole" connections") are hyper-parameters that can be optimized. For example, stacking several recurrent layers by feeding the output of each LSTM neuron in one recurrent layer as input into an adjacent recurrent layer, or increasing the number of neurons and hidden units, enable more complex or hierarchical representations at the risk of over-fitting. Furthermore, the number of output neurons, which represents the number of functions to be simultaneously considered (i.e. multiplex), can be varied. In this work, a single set of bi-directional recurrent layers was utilized for in-class predictions, and up to three sets were used for training toward out-of-class predictions. As protein sequences vary widely in length, the number of LSTM neurons in the recurrent layer was capped, typically at 333 representing a maximum of 333 amino acids sequence or around 1 kilo-base of DNA. For proteins smaller than 333 residues the sequence was pre-padded with 0's up to a 333-digit sequence, where the digits 1 to 20 represents the 20 canonical amino acids. For functions with mostly large proteins such as the CRISPR-associated nucleases, up to 800 N-terminal amino acids were input for training, and subsequently the same RNN model was trained on up to 800 C-terminal amino acids. There are 128 hidden units in each LSTM neuron (i.e. the hidden state is represented by a 128 element vector). A high dimensional hidden state vector can encode more information to represent more complex function-related sequence features. This can be an advantage compared to some sequential models (e.g. hidden Markov model) with limited number of internal states at the expense of interpretability. Additionally, the multiple nonlinear operators of the LSTM (e.g. activation functions) allow complex updating of the hidden state memory. Adding to this flexibility, the probability of "Dropout", the random severing of connections between layers, was consistently set to 0.5. Unlike previous artificial neural network based methods, the LSTM model here does not limit itself to learning short "profiles" or motifs of pre-defined length[3–5] (e.g. 21 amino acid window[3]) but instead learns from the entire sequence up to a maximum length (e.g. 333, 500 or 800 from each terminus) in order to capture potentially long-range patterns.

*In-class Model Training and Validation*

For training, protein amino-acid sequences were obtained from the UniProt database and directly used as inputs into the neural network without any feature extraction. For prediction of a particular protein function, the positive class contains all sequences that match the function in UniProt by keyword. The negative class contains all non-matching sequences in SwissProt (the manually reviewed database within UniProt with currently around 550K sequences). Of the combined dataset, 80% was randomly selected and employed as "train-set" for training the neural network, and the remaining 20% was used as the "test-set" to evaluate the trained model's performance on yet-unseen dataset. As the negative class generally greatly outnumbers the positive class, it was divided into 4 or more chunks to train against the positive class sequentially for class-balance. Each chunk of the negative set combined with the positive set was trained for at least 5 epochs (i.e. passes over the entire dataset) during which the categorical entropy of the predicted output compared against the expected output was minimized via the ADAM optimizer[6] with the mini-batch sampling size set to 64. Ten percent of the total data within the training dataset was used to monitor the network losses and changes in prediction accuracy during each training step. Furthermore after each chunk had been trained, the prediction performance on the "test-set" was evaluated to calculate the accuracy, precision, recall and F1 (F-measure) for the positive and negative classes. The "test-set" data was initially selected and mixed at random without applying class-balance in order to mimic real-life operations when the positive class is heavily under-represented.

Four functional classes were picked to test the performance of the RNN predictive model: iron sequestering proteins (class "Ferritin"), cytochrome P450 proteins

(class "P450"), serine and cysteine proteases (class "Protease") and G-protein coupled receptors (class "GPCR"). Iron sequestration is crucial for cells to maintain iron homeostasis and protect against ROS generation from iron-catalyzed Fenton reactions[7]. Currently well-known iron sequesters across domains of life are ferritins and dps (DNA-binding protein from starved cells) proteins which form protein cages in cells that sequester and mineralize iron into inorganic nanoparticles. In addition to detoxification, the iron oxide nanoparticles synthesized could potentially be utilized towards noninvasive applications in biology such as a reporter or contrast agent for magnetic resonance imaging[7]. The iron sequestration and magnetic properties of the proteins can be experimentally validated using cellular assays[8]. P450 proteins are also ubiquitous across kingdoms of life and are enzymes that act on a variety of substrates carrying out important tasks including detoxifying drugs in humans. G-protein coupled receptors are important trans-membrane proteins for cellular signal transduction and are targets for many drugs. Lastly, serine and cysteine proteases cleave peptide bonds in proteins to break them down and represent a prime example of molecular-scale convergent evolution, where different organisms independently evolved the "catalytic triad" for performing the peptide cleavage function with otherwise little homology at the overall protein sequence level.

High performance of predictions on the randomly left-out "test-set" data not seen by the model during training was obtained for all four classes of protein functions (Figure 2a). Even though accuracy is nearly 100% for all predictors, it is not the most informative measure as the negative class of proteins not possessing a particular function vastly outnumbers the positive class and a predictor could achieve high accuracy by simply only predicting negatives. But despite such challenge of "finding needle in a haystack", all functional predictors were able to achieve close to unity precision and recall in identifying the correct sequences from the "test-set", with F1 measure close to unity. The receiver operating characteristic (ROC) plots for the True Positive Rate (sensitivity) versus False Positive Rate as a function of the classification threshold (between 0 and 1) and their Area Under the Curve (AUC) close to unity also demonstrates the model's ability to make strong discrimination of the positive class distribution from that of the negative class in the tested dataset (Figure 2b). However, it is important to note that these metrics do not readily apply to prediction on arbitrary datasets, particularly large databases where class imbalance (ratio of negatives to positives) is extreme due to the negligible fraction of total proteins that have one specific function, and RNN performance may be negatively impacted. Also very low false positive rate (e.g. 1E-6) would be needed to avoid large number of false positives when searching a large database (e.g. 54 million sequences in UniRef100). Lastly, as anticipated, the prediction performance in precision and recall decreases as the cutoff length of the input sequence or equivalently the depth of the bi-directional recurrent layer was decreased as demonstrated for the "Ferritin" class (Figure 2c, d), even though reducing neural network depth increases training speed. Allowing input sequence length greater than 333 amino acids significantly increases processing and memory requirements without yielding noticeable increases in prediction performance for the four protein functions of interest.

*Database Search and Prediction*

The trained and performance-validated models were used to predict whether a new sequence without assigned function could possess a potential function. Currently, state-of-art tools for remote homology search include PSI-Blast, Delta-Blast and in particular jackhmmer (part of HMMER[2]) which utilizes Hidden Markov Models. For comparison, HMMER and the RNN models were run on the same comprehensive UniRef100 sequence database containing numerous uncharacterized or unannotated proteins sequences. For each of the four functional classes (Ferritin, P450, Protease, GPCR), a representative or important member was used (FTNA_ECOLI, CP21A_HUMAN, SEPR_HUMAN, FFAR2_HUMAN), respectively, as initial seed for iterative HMMER (jackhmmer) search on the UniRef100 database, and at least 5 iterations were run with a reporting cutoff threshold of e-value E=10.0 (default). Separately, the trained RNN models also predicted thousands of new hits from the UniRef100 database for each function (Figure 3a, "Predict") *in addition to* the thousands of sequences that were used for training each model. Upon comparing the lists of outputs from HMMER to the RNN models discounting the already-annotated sequences used for training, there were still thousands of new, unique sequences predicted by the RNN model that were not shared by the HMMER output (Figure 3a, "Unique"). As a check, the majority of additional sequences predicted by the RNN model have some identification of the correct family or gene ontology in a public database obtained by other sequence or structure homology detection techniques (Figure 3b). However, there is a further subset of the predicted sequences that are unannotated and uncharacterized in UniProt (Figure 3a, "No Annot."). For the "Ferritin" class, the "No Annot." sequences predicted by the RNN show numerous lineages after multiple sequence alignment by Clustal Omega (using EMBL-EBI server), suggesting a set of diverse, dissimilar sequences not sharing obvious sequence patterns identifiable by alignment (Figure 3c). The statistics of the domains of origin for the predicted proteins reveal certain domain biases for function, such as bacteria for "Ferritin" class or eukaryote for P450 and GPCR, as expected (Figure 3d). Similar biases could be seen for the "Ferritin" and "P450" classes in the taxonomy of the organisms of origin for the predicted proteins (Figure 3e).

*Experimental Validation of Predicted Function*

To validate the functional prediction by the RNN model of sequences without characterization or annotation in UniProt, I experimentally characterized the iron sequestration properties of ten "unique" candidates predicted by the RNN model for iron sequestration proteins. The ten sequences were selected from diverse domains of life and vary widely in their amino acid lengths and composition (Figure 4a). The candidates were named after their biological contexts. Homology search with these sequences as seeds using popular bioinformatics tools such as BLAST and jackhmmer using their web servers on the latest protein databases (NCBI nr, Reference Proteomes) yielded mostly proteins of unknown (only "predicted" or "hypothetical") and uncharacterized function. However, some functional homologues were identified. For the "fungi" candidate, both web-based BLAST and jackhmmer were able to detect "ferritin-like" homologues, corroborating the RNN prediction. On the other hand, candidates "human", "mouse", "potato", "cyano", "gut" or their BLAST/jackhmmer homologues showed few entry names suggestive of other functions such as "Alternative protein NCAM1 (neural cell adhesion molecule)" for the "human" candidate and "poly-homeotic like protein" for

"mouse" candidate. This could have new implications for the biological activity, particularly of iron sequestration, for these uncharacterized sequences. The remaining candidates "lancelet", "virus", "algae" and "archaea" yielded no hint of protein function. The DNA sequences encoding all 10 uncharacterized proteins were codon optimized, synthesized, and cloned into vectors in *E. coli* cells and expressed highly using a rhamnose-inducible, high copy number vector. The *E. coli* cells simultaneously contain a fluorescent, genetic iron sensor based on the *E. coli* fiu promoter that has been validated to detect intracellular iron depletion (Chapter 3). Using calibration by iron chelator bipyridine, the fluorescence values could be converted to equivalent intracellular free iron concentrations. After induction of recombinant protein expression during exponential growth phase followed by overnight growth to saturation in LB media supplemented with 100µM Fe (II) sulfate, the cells were characterized for their fluorescence by the green fluorescent protein (GFP) reporter. All ten proteins showed statistically significant increases in fluorescence, or equivalently decreases in cellular free iron concentrations upon protein expression relative to no expression/induction (P value < 0.05 by two-tailed Student's t-test) (Figure 4b). However, the protein derived from "potato" did not dramatically change the concentrations compared to the others. To determine the proteins' ability to not only bind and sequester iron but also to bio-mineralize similar to the ferritins and dps proteins, I measured the retention level of the protein-expressing cells in high-gradient magnetic separation columns, as iron based minerals could increase magnetic moment of the cells. Some of the proteins tested, particularly "algae", "human" and "archaea", demonstrated increased magnetic retention compared to the uninduced control (Figure 4c). The expression of some of these proteins including "algae", "archaea", "virus", and the non-sequestering "potato" were clearly observed by SDS-PAGE gel (Figure 4d). The inability to observe bands for candidates "human" and "mouse" may be due to their very low molecular weight (predicted <10kD). Furthermore, the impact of mutations to the predicted sequences on the desired iron-sequestering function could be analyzed using the same trained RNN model *in silico* in the manner of "saturation mutagenesis" where residue position of a sequence is mutated to every other base. The resulting impacts are illustrated in heat-maps with the residue positions along the sequence along the horizontal axis and the 20 canonical amino acids along the vertical axis. The negative impacts are illustrated as red and positive impacts as green. In this manner, residues "conserved" for function are easily identified by the "red columns" (Figure 4e). Furthermore, a "red row" at proline illustrates the potential helix-breaking and structure-disrupting property of proline, a chemical property that the RNN model has learned only from sequence information without a priori chemical knowledge. Further experimental testing of such mutations could enable further validation and optimization of the RNN model. Lastly, homology modeling of some of the predicted candidates using I-TASSER[9], the top structure prediction method in the CASP competition in 2012 and 2014, reveals diverse structures. Therefore, the model is not predictive of a particular protein fold or structure but other sequence-based features associated with function.

*Comparison of RNN to Other Machine Learning Methods*

For machine-learning benchmark, the performance of the RNN model was compared against other popular machine learning classification models, particularly logistic regression and random forest which are known for speed, robustness and often good predictability. Furthermore, both algorithms are capable of modelling nonlinear relationships as would be expected between protein sequences and functions that would not be accurately captured by other fast machine learning methods such as linear regression. For all of these models, a set of features or independent variables are required. Using the same dataset for each of the four functional classes, 51 ProtParam features (Table S1) were extracted or calculated for each sequence and vectorized. These features include simple amino acid composition and length as well as biochemically relevant properties such as isoelectric point, molecular weight, stability index, hydrophobicity and grand average of hydropathicity (gravy). The logistic regression and random forest models were each trained using "grid-search" over a range of values for their model hyper-parameters, such as alpha for logistic regression, and the parameter values that produced the best prediction results were selected. Comparing the "in-class" prediction performance on the four functional classes by all the machine learning methods, logistic regression was by far the fastest to train but also the least predictive (Figure 5). While random forest was slower, it achieved much better performance but still outclassed by the near perfect performance of the RNN model on the same dataset. Nonetheless, the "feature importance" of random forest models calculated for the four predictors on the 51 features reveals different biases toward different functional classes (Figure 5). The RNN model could not be simply interpreted based on these predefined features, but their best-in-class performance without "feature engineering", like in other successful deep-learning applications, demonstrate their potential to represent and capture nontrivial and difficult-to-quantify patterns or relationship between sequence information and protein function.

*Out-of-class Training and Prediction*

Lastly "out-of-class" prediction performance was tested, whereby the RNN models were trained on sequences from certain protein families and tested on other functionally homologous but phylogenetically distinct families. One drawback of the random splitting of UniProt dataset into train and test sets employed so far is that the two sets could contain highly similar or even identical sequences that represent homologous proteins from closely related species. Furthermore, the ability to discover proteins with homologous function that are distant in evolution from what are already known could be valuable both for studying sequence evolution as well as mining for novel proteins for particular applications like genome-editing. Here I conducted "out-of-class" prediction test on three functions, "Genome-Edit", "Ferritin", and "P450". The negative set for both training and testing was again the reviewed SwissProt database excluding members containing function of interest. For the "Genome-Edit" function, RNN was trained on the InterPro Cas9 family of proteins (IPR028629, 1201 sequences) as positive set and tested on the InterPro Cpf1 family of proteins (IPR027620, 55 sequences)[10]. Both Cas9 and Cpf1 are guided nucleases associated with the CRISPR locus[11,12,13]. Cpf1 was discovered more recently and confer benefits such as not requiring a "tracr-DNA" for targeting and potentially higher specificity. Due to the scarcity of the positive training set (Cas9 family) relative to the set of negatives (>550,000 in SwissProt outside of Cas9 and Cpf1 family), the negative set was divided into 100 chunks and sequentially trained with the same positive set (Cas9 family). Such class-balancing or under-sampling during training was not applied during testing on the Cpf1 to more closely simulate the naturally small fraction of positives in a database. For the "Ferritin" function, RNN was trained on the InterPro non-haem ferritin family (IPR001519) along with *either* the haem-containing bacterioferritin family bfr (IPR002024) *or* the DNA-binding protein dps family (IPR002177) as positives, and tested on the remaining un-trained family. The dps differs from the ferritins or bacterioferritins prominently in assembling a cage of 12 rather than 24 monomers. The "P450" function is represented by 6 different sequence clusters/families in InterPro: B-class (IPR002397), E-class-CYP24A-mitochondrial (IPR002949), E-class-group-I (IPR002401), E-class-group-II (IPR002402), E-class-group-IV (IPR002403) and mitochondrial (IPR002399). Either the B-class (31205 sequences) or E-class-group-II (2314 sequences) was treated as the test-set, with training of RNN using the combination of the other families as positives. Taking into account the different length distributions of the protein families, the maximum recurrent depth (i.e. sequence-length) was capped at 333 for "Ferritin", 500 for P450, and 800 for "Genome-Edit". To remove possible false positives in the training sets, sequences shorter than 10 amino acids or longer than 1000 amino acids for "Ferritin", "P450" functions, or 2000 amino acids for "Genome-Edit", were filtered out before training. As the "Genome-Edit" Cas9 or Cpf1 enzyme sequences are typically over 1000 amino acids long, the RNN was trained scanning over up to the first 800 amino acids from the N-terminus and subsequently from the C-terminus. Overall, prediction performance varied more substantially among the out-of-class predictors compared to the previous random, in-class prediction performance (Table 1). Decent detection sensitivities were achieved with the left-out P450 families and for detecting bfr after training on non-haem ferritins and dps. However, sensitivity/recall was low (0.13) for detection of the 12-member caged dps from RNNs trained only on the 24-member caged non-haem ferritins and bfr. Tripling the number of recurrent layers by feeding the output sequence of one layer as input into the next, which produced a slower but deeper model with potential to encode more complex sequence patterns, increased sensitivity for detecting dps from 0.13 to 0.36 without decreasing precision. Lastly, prediction performance on Cpf1 from an RNN trained on Cas9 yielded sensitivity/recall of 0.59 after training on both N and C terminal residues (up to 800 amino acids) and averaging the prediction probabilities of processing the sequence from its two termini for final classification. Interestingly, classifying using predicted probabilities for only the N- or C-terminal residues (up to 800) significantly decreased precision (i.e. increased false positives), suggesting that multiple features along the entire sequence length (e.g. the binding and nuclease domains) may be required toward accomplishing the "Genome-Edit" function and that many other proteins may exist with only a subset of those features.

**Discussion**

In summary, this study has shown that recurrent neural network (RNN) based on LSTM can be trained to classify certain protein functions with high level of accuracy from input amino acid sequences alone. Experimental validation of the predicted iron sequestering or mineralizing proteins including some currently not easily identified by other bioinformatics methods confirm the accuracy and utility of the model for prediction.

Compared against popular sequence prediction and analysis tools such as BLAST and HMMER, the RNN model currently has several potential benefits but also limitations. One important benefit is the potential to capture obscure sequence-function relationships, allowing predictions of very remote homologies. Unlike most sequence search tools, RNN models do not explicitly rely on sequence alignments or heuristic scoring functions or similarity measures. The memory or internal state of the LSTM neuron processing entire protein sequences, unlike other machine learning methods that employ short, pre-defined motif windows[3–5], allows selective retention of important sequence features across long distances[14]. For instance, residues that make up an active site of an enzyme may be separated by large gaps in the protein sequence, but are in proximity of each other in 3-dimensional space. Despite much advances in recent years, the folded structure of proteins still cannot be reliably predicted from their primary amino acid sequences, which limits the prediction of protein function most often highly related to the structure. In this work, four important functional classes were selected which includes as their members proteins across domains of life that share little homology, or have converged upon the same function without common evolutionary origin as in the catalytic triad of the proteases. The ability of the RNN model to accurately make predictions for all of these functional class from only primary sequence without structural information suggests that the RNN could represent complex patterns in the protein sequence that encode for function. However, it is important to note that the "in-class" performance measures obtained from testing on randomly selected sequences from a small and predominantly reviewed dataset (fewer than 1 million sequences) may not hold for testing on arbitrary databases (e.g. UniRef100 with 54 million sequences). In the larger databases, the proportion of members with particular function (i.e. the positive class) can be extremely small. As a result, very high performance is demanded, with false positive rate approaching zero to avoid large number of false positive predictions. The "in-class" performance, though respectable, will require calibration on the same test databases for comparisons against current state-of-art (e.g. BLAST). Furthermore, the "in-class" predictors' performance may partially benefit from the high similarity or possible redundancy of sequences representing homologous proteins in closely related species randomly partitioned into the training and testing sets. The "out-of-class" predictors tested on phylogenetically distinct families showed lower performance as expected. Therefore, while the RNN models can be sensitive toward new protein families with functional homology, further optimizations are necessary to improve their sensitivity and selectivity particularly for this difficult task of discovering new protein families with related functions in the large and growing sequence databases.

  As a "deep-learning" model, RNN with LSTM has found success in several domains related to sequence learning, particularly language recognition and modelling, that surpassed the performance of other machine learning models particularly for learning directly from raw data[15]. However, a current limitation of using RNN with particularly deep layers (e.g. long sequences) is the training and processing speed. This is mainly because of the large number of variables in a deep neural network model which requires training with large datasets and many operations on large matrices in the iterative optimization steps using the relatively slow gradient based, backpropagation techniques. Building Position Specific Scoring Matrices (PSSM) for PSI-BLAST or hidden Markov models for HMMER as well as searching against those models can be performed faster currently on the public servers.

Besides currently limited computing power, another limitation at the present is the data itself. While there is abundant data for accurate training for the functions of iron-mineralizing proteins, cytochrome P450s, proteases and GPCRs, there are some functions of interest that at the present do not yet have sufficient data size to produce highly predictive models. For example, in the last few years there has been exploding amount of interest and applications of oligonucleotide-targeted nucleases for genome-editing across a variety of systems. The CRISPR (Clustered Regularly Interspaced Short Palindromic Repeats) system originated from bacteria *Streptococcus pyogenes* has been particularly successful in efficient genome editing across a variety of cell types including human cell lines[12,13]. And in recent years new systems of similar function are continuously discovered via bioinformatics techniques for remote-homology prediction such as PSI-BLAST[1]. It is of great interest to discover the whole diversity of oligonucleotide-targeted nucleases for future enhancement of genome editing applications. While there may already be some that are homologous to the known CRISPR systems by sequence or structure, there are potentially more in Nature with more remote homology not detectable by the PSI-BLAST or HMMER. The deep learning approach here employing RNN has the potential to detect those remote candidates. However, the main challenge currently is the limited amount of public data for creating the training set, as fewer than 10,000 CRISPR/Cas9 like nucleases have been identified. Additionally, unlike the iron-mineralizing ferritins or P450s, these guided nucleases so far identified are mostly large proteins with relatively long sequences of more than 1000 amino acids. Long sequences have been particularly challenging for RNN training due to the exploding or vanishing gradient issue with back-propagation. The use of LSTM neurons allowing selective retention and forgetting of information has ameliorated the issue, but training very long sequences would require significantly more computational processing power and memory. Given significantly more computational resources and time, results here have shown that deeper RNN models could be trained on the currently available dataset to make reasonable predictions (Table 1). However, both sensitivity and selectivity could be optimized with training on the growing volume of experimental data in order to more accurately and precisely discover new functional candidates or protein families and demonstrate utility and power of the RNN predictors over the current state-of-art (e.g. PSI-BLAST).

     Despite current limitations in speed and data availability for certain functional prediction applications, RNN-based deep-learning models have the potential to overcome these obstacles quickly in the coming years to become more widely applicable enabled by three trends. On the speed side, both the cost and performance of computing are improving rapidly, particularly due to the design and deployment of highly-parallelized processing architectures (e.g. graphic computing units) that are

particularly well suited and have been increasingly dedicated toward training deep neural networks. On the data side, increasingly large volumes of data are collected from automated, high-throughput experimentation. In the field of synthetic biology, first the cost of sequencing and now of synthesis of DNA has been decreasing dramatically. Large throughput sequencing, particularly of hard-to-culture environmental samples in metagenomics, has rapidly increased the database of sequences available for mining new proteins and new functions. Meanwhile, the accessibility of DNA synthesis has made it possible to quickly test new sequences of interest in relevant biological contexts and obtain valuable data such as those related to protein functions. As more validation data become available, the deep-learning model can be further trained to become more powerful at predicting desired functions. As the RNN is agnostic to the specific biological nature of the sequence, it can be potentially useful for analyzing other biological sequences besides amino-acids (e.g. RNA). Furthermore, as RNN can be a generative model, it can be trained on proteins of a particular functional class with an auto-encoder and use the decoder to write new protein sequences that may possess that function. This is currently done for translating human languages[16,17] due to the abundance of data. It may be foreseeably applied to protein sequences in the future as the amount of data increases, but it will be significantly more challenging due to requiring the RNN model to learn and remember not only sufficient patterns for classification of certain functions but also everything else that makes a functional protein, as often even few mutations unrelated to a particular function could cause proteins to misfold. At the very least, much deeper RNN models (with numerous stacked recurrent layers) and large hidden state vectors that are capable of storing more information, along with ample training dataset for not only particular function but also for other essential aspects such as proper protein folding, will be necessary to accomplish *de novo* protein "writing". Lastly on the theoretical side, the convergence of artificial neural networks (ANN) research with the field of neuroscience where it first drew its inspiration could lead to potentially better model or computing architectures that improve both the speed and accuracy of the artificial recurrent neural-net based predictors.

**Materials and Methods**

**Computational Modelling**

All computational models were written in Python and processed on the Harvard Odyssey computing cluster at Harvard University using a combination of CPU and GPU computing nodes. The recurrent neural network models were built upon the Google Tensorflow backend. The logistic regression and random forest models were built using the Python scikit-learn packages. HMMER v3.1b1 (jackhmmer tool) was deployed and executed also on the Odyssey cluster. Protein sequence and function data were obtained directly from the UniProt databases (www.uniprot.org)

For each LSTM Neuron in the RNN, its input "i", output "o", gate "g", forget "f", cell state "c" and hidden state "h" values at time "t" are determined by the following equations[14,18]:

$$i_t = \sigma(D(x_t)W_{xi} + h_{t-1}W_{hi} + b_i)$$

$$f_t = \sigma(D(x_t)W_{xf} + h_{t-1}W_{hf} + b_f)$$

$$g_t = \tanh(D(x_tW_{xg}) + h_{t-1}W_{hg} + b_g)$$

$$c_t = f_t \circ c_{t-1} + i_t \circ g_t$$

$$o_t = \sigma(D(x_t)W_{xo} + h_{t-1}W_{ho} + b_o)$$

$$h_t = o_t \circ \tanh(c_t)$$

$$\sigma(z) = \frac{1}{1 + \exp(-z)}$$

, where *W* represents weight matrix, *b* represents constant bias, *D* represents dropout (sets value to zero with probability *p*, *p*=0 in this study), ∘ represents element-wise multiplication (Hadamard product), and *tanh* represents hyperbolic tangent function.

For evaluation of machine learning performance, the metrics are computed from the number of True Positives (TP), True Negatives (TN), False Positives (FP) and False Negatives (FN) as follows:

Accuracy = (TP + TN) / (TP + TN + FP + FN)

Precision = TP / (TP + FP)

Recall = TP / (TP + FN)

F1 = 2 x Precision x Recall / (Precision + Recall)

True Positive Rate (Sensitivity) = TP / (TP + FN) = Recall

False Positive Rate = FP / (FP + TN)

The Receiver Operating Characteristic (ROC) is plotted for the True Positive Rate against the False Positive Rate as classification threshold is varied.

The performances reported in the main text and figures consider the proteins containing a particular function, the minority class, as "Positive" whereas the rest are "Negative".

**BLAST and HMMER search of experimentally validated RNN predictions**

Default setting were used for NCBI BLAST and EMBL-EBI jackhmmer on their web-servers for searching the ten RNN predictions that were experimentally validated for possible functional homologs. Specifically for NCBI BLAST, the NCBI non-redundant protein sequences database was used for blastp. For jackhmmer run on the EMBL-EBI server, the Reference Proteomes was used, and the Cut-Off thresholds were set at default values such that Significance E-values was 0.01 for sequence and 0.03 for hit, while the Report E-values were 1 for both Sequence and Hit. Jackhmmer was iterated until convergence.

**Construction of expression vectors for predicted protein candidates**

Candidate genes for experimental validation were first synthesized as gene-Blocks (gBlocks) according to their sequences. The N-terminal six methionine repeat sequence of "human" was synthesized with only the last methionine due to DNA synthesis difficulty of ATG repeats and the possibility of product from translational start at the last methionine. The gBlocks were then cloned into a high copy-number plasmid (pUC origin of replication) with rhamnose inducible promoter (*rhaP$_{BAD}$*, with native *E. coli* transcription factors RhaS and RhaR) and kanamycin resistance cassette via Gibson Assembly. The DNA plasmid was verified by Sanger Sequencing (Genewiz) and transformed into *E. coli* BW25113 cells via electroporation. Protein expression was induced in cells by adding rhamnose to cell culture (maximum 0.2%) during log-phase growth (OD600~0.4). DNA sequences of the most relevant genes and constructs can be found in Table S2 in Appendix C.

**Iron level characterization by genetic sensor**

For the genetic iron sensor, the *E. coli* fiu promoter was cloned along with a super-folder GFP (sfGFP) reporter via Gibson Assembly into a low copy (p15A origin), chloramphenicol-resistance plasmid compatible with the ferritin-expressing plasmid. Iron levels were measured for cells containing the protein-expression and iron sensor plasmids by taking the GFP fluorescence of the culture of cells (488nm excitation by laser, 512nm emission) in 96-well plate format using the BioTek NEO plate-reader. For calibration, known concentrations of iron sequesterer bi-pyridine were added to cell cultures. The fluorescence measured were normalized to culture density by dividing by OD600 measured by the same plate-reader. The increase in normalized fluorescence of the cells was plotted against the increase in bipyridine (or consequent decrease in free iron) and modeled to determine the conversion between fluorescence reading and free iron concentration[8].

**Magnetic Column Retention characterization**

A high-gradient magnetic column (Miltenyi LD columns) was sandwiched between two neodymium permanent magnets (K&J Magnetics Inc., BX8C4-N52) to create high magnetic field gradients inside the column. The column was first wetted by passage of 2ml of PBS 1X buffer. Then 500µl of cells re-suspended in PBS 1X buffer were added and flowed through by gravity into the elution tube, followed by addition of 3ml of PBS 1X buffer to wash through any unbound cells into the elution tube. Once dry, the column was removed from the magnets, and 3ml of PBS buffer was pushed through the column to extract the magnetically retained cells into a separate retention tube. Measuring OD600 of the elution and retention tubes allow estimation of cell counts and the percentage of total cells retained by the magnetic column.

**SDS gel analysis of protein expression**

*E. coli* cells were re-suspended in SDS Buffer (NuPAGE LDS Buffer) with reducing agent, followed by two cycles of boiling at 95°C for 5 minutes and vigorous vortex to lyse cells and denature proteins. The lysate was centrifuged to pellet cell debris, and the protein suspension was diluted and added to NuPAGE 4-12% Bis-Tris gel with MES buffer. Empty lanes in the gel were filled with equal volume of SDS buffer. After running at 200V for 35 minutes, the gel was removed and stained with Coomassie Orange dye for one hour and subsequently imaged for dye fluorescence on a Typhoon Imager.

**Acknowledgement**

The author would like to acknowledge Professors Sean Eddy, Debora Marks and Pamela Silver at Harvard University for helpful discussions and feedback. The author would like to acknowledge the Harvard Odyssey Computing Cluster for providing the computational resources for this work.

**Conflict of Interest**

The author declares no conflict of interest for this work.

**Figures**

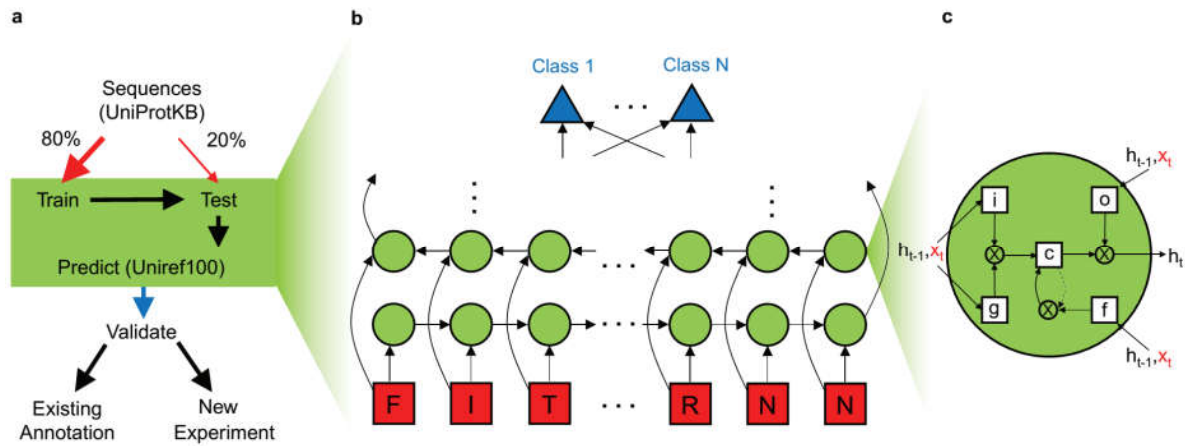

**Figure 1 Machine-learning model for protein function prediction (a)** workflow of the prediction model consists first feeding sequence dataset with known functional annotations. After training the machine-learning Recurrent Neural Network (RNN) model with 80% of the sequences chosen randomly, the last 20% of yet unseen sequences are fed to test the model prediction performance. Alternatively, the model can be tested on sequences of protein families with homologous function but distinct phylogeny from the training set (e.g. "out of class"). The tested model is used to scan and predict all proteins (including unannotated) in the UniRef100 database. The positive predictions are validated either by existing annotation (e.g. in UniProt) or experiment **(b)** The RNN model consists of arbitrary sets of forward and reverse layers of long-short term memory (LSTM) neurons taking only the amino acid letters from the sequence as input (red). The final output of the recurrent layers are combined into a fully connected layer for functional classification (blue) **(c)** Each LSTM neuron contains gates for input "i", output "o", gate "g", and forget "f", which update along with the new input the cell state "c" and hidden state "h" to encode relevant sequence patterns.

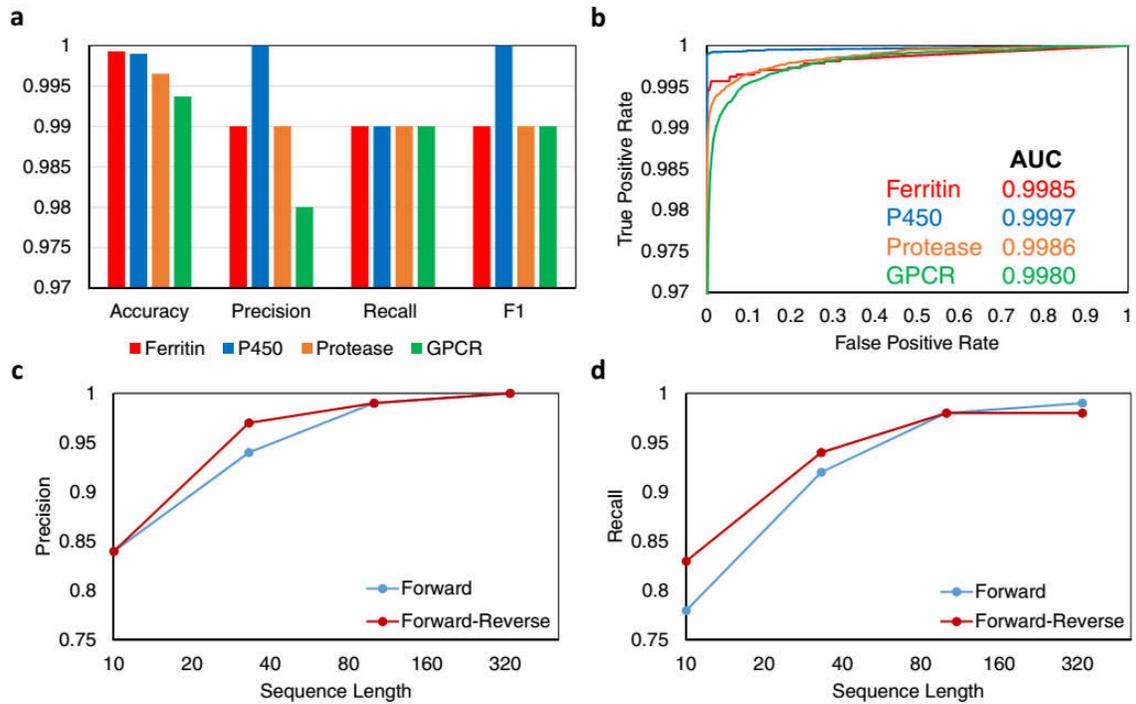

**Figure 2 RNN model achieves high prediction performance on randomly left-out testing data (a)** High prediction performance is achieved for all four tested classes: iron-sequestering (Ferritin), cytochrome P450, protease (serine and cysteine) and G-protein coupled receptor (GPCR). **(b)** Receiver-operating characteristic (ROC) of the four separate models demonstrate high Area-Under-the-Curve (AUC). For the "Ferritin" class, prediction precision **(c)** and recall **(d)** both improve to close to unity as the length of amino acid sequence shown the network increases, saturating around 333 letters.

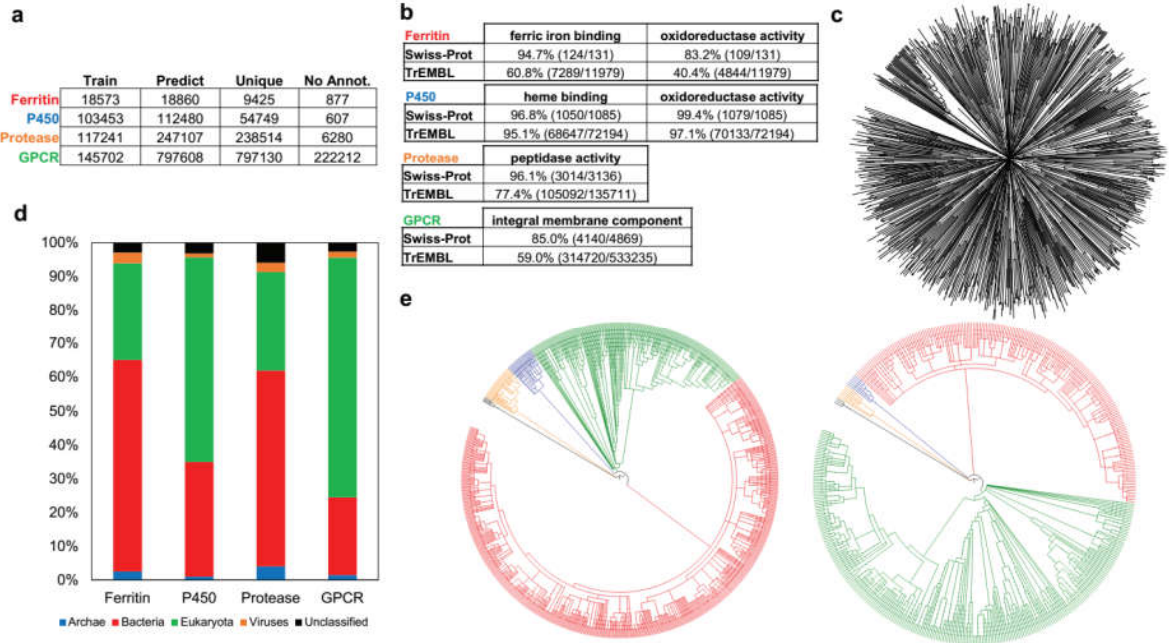

**Figure 3 Trained RNN model predicts new annotations (a)** Table listing for each function, the number of sequences used for training the RNN model, the number of *additional* sequences it predicted as positive in the UniRef100 database ("Predict"), the number of sequences not included in output of jackhmmer (iterative HMMER search) using representative starting sequence ("Unique"), and the number of sequences without any function or family annotation on UniProt and linked databases (Gene3D, InterPro, PROSITE, Pfam, SUPFAM) ("No Annot."). **(b)** Among the "Predict" sequences, high percentages agree with manually curated Swiss-Prot annotation for expected gene ontology of each class. Agreement is worse for the automatic annotations in TrEMBL database particularly for "Ferritin" and "GPCR" functions. **(c)** Clustal Omega multiple sequence alignment of the "No Annot." sequences for "Ferritin" function shows diverse lineages. **(d)** Taxonomy of "Predict" proteins reveals expected bias for functional class. **(d)** Taxonomy of the organism of origin for the "Predict" proteins for "Ferritin" (*left*) showing greater species diversity among bacteria (*red*) and "P450" (*right*) showing greater diversity among eukaryotic species (*green*).

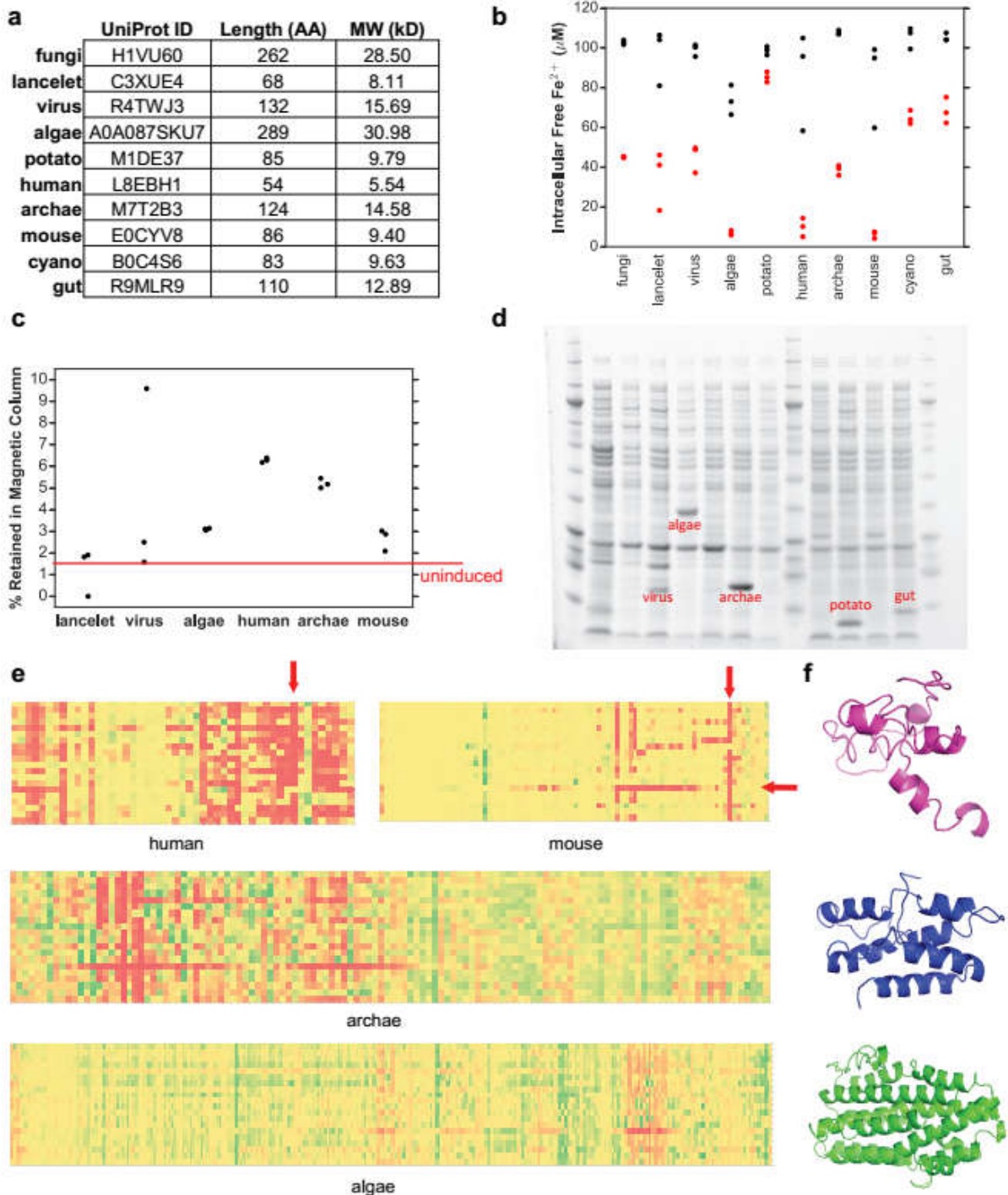

**Figure 4 Experimental validation of predicted iron sequestering proteins (a)** List of ten proteins picked from diverse biological contexts without annotation in UniProt predicted by RNN model to contain "Ferritin" like function **(b)** after cloning and expressing the proteins in *E. coli* with genetic iron sensor, the majority of the tested proteins demonstrated decreased cellular iron particularly for "algae", "human", "archaea" and "mouse". (P value < 0.05 by two-tailed Student's t-test. Three biological replicates in one experiment. Iron sensor functionality has been verified with controls

and other protein sequences[8], replicated more than three times in lab.) **(c)** Several candidates also gave rise to increased cellular magnetism (magnetic column retention) due to possible iron biomineralization compared to uninduced cells (3 biological replicates in one experiment.) **(d)** Bands for over-expressed proteins could be clearly observed for "virus", "algae", "archaea" and "potato" (did not demonstrate significant iron sequestration or magnetism) **(e)** *in silico* "saturation mutagenesis" of selected sequences using RNN model to predict effects of mutations on desired function (red=bad, yellow=neural, green=good), with residue position along horizontal axis and the 20 canonical amino acids along vertical axis. RNN model identifies key positions conserved for function (e.g. vertical arrow), and also the potentially structure-breaking mutations by mutation to proline (horizontal arrow) **(f)** structural homology models of protein candidates "mouse" (*top*), "archaea" (*middle*) and "algae" (*bottom*) using I-TASSER server (the top method in the recent CASP 2012, 2014 protein structure prediction competitions), showing diverse predicted structures.

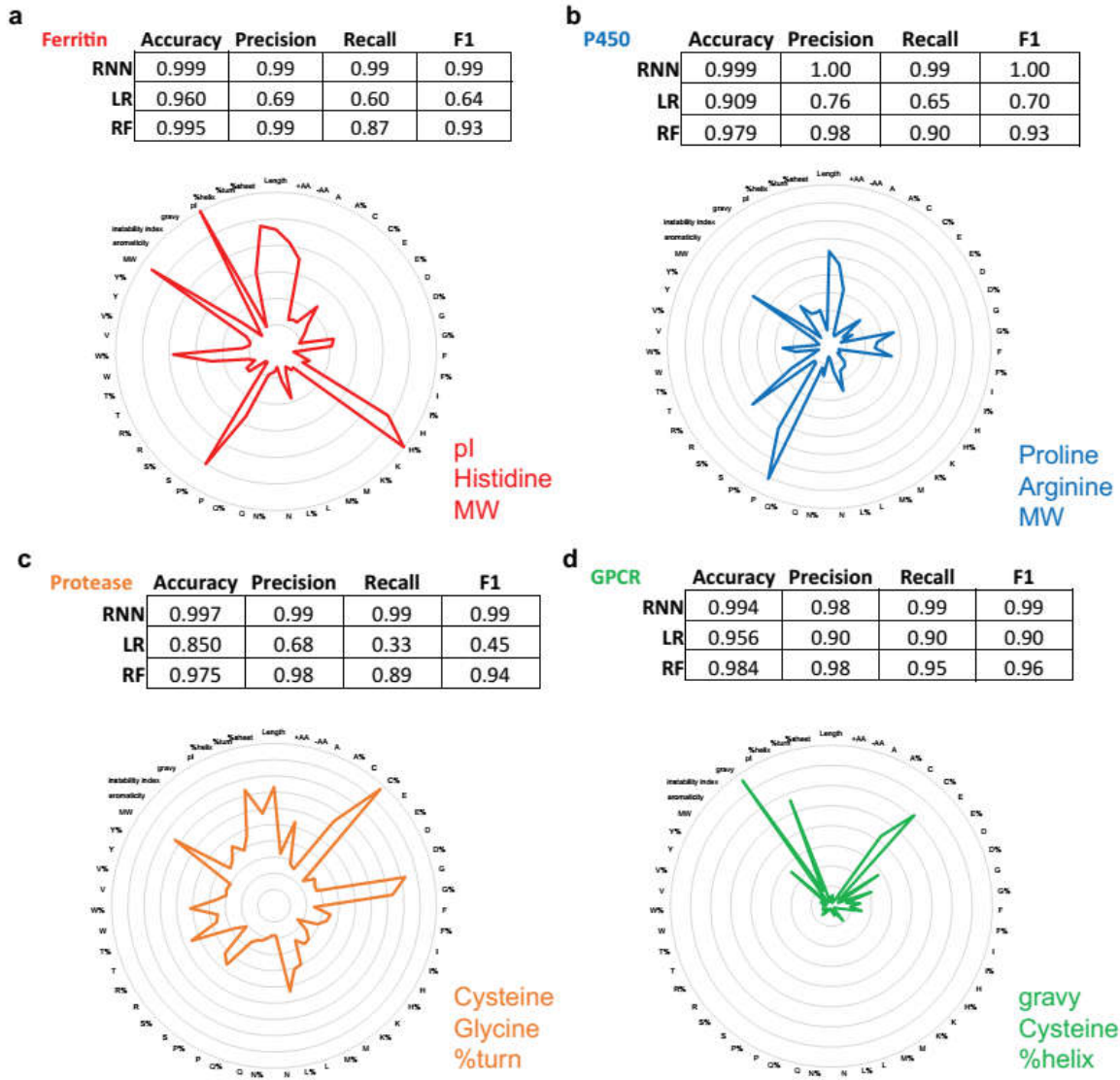

**Figure 5 Performance benchmark with other machine-learning classifiers** For each of the functions "Ferritin" **(a)**, "P450" **(b)**, "Protease" **(c)** and "GPCR" **(d)**, separate logistic regression (LR) or random forest (RF) models were trained on the same input sequence set with 51 sequence-derived ProtParam features optimized by grid-search of hyper-parameters and 5-fold cross-validation and used for prediction on 20% of randomly left-out unseen dataset. The RNN model outperforms in accuracy, precision, recall and F1. To assist understanding the learning of the models, "Feature importances" of the RF models, which achieved relatively high performance, are shown in radial-plots. The three most important features for RF predictions are listed for each function (gravy: grand average of hydropathicity).

|  | precision | recall | F1 |
|---|---|---|---|
| **Ferritin-bfr** | 0.98 | 0.59 | 0.74 |
| **Ferritin-dps** | 0.96 | 0.13 | 0.22 |
| **Ferritin-dps_3X** | 0.99 | 0.36 | 0.52 |
| **P450-B** | 0.93 | 0.81 | 0.87 |
| **P450-E_II** | 0.61 | 0.91 | 0.73 |
| **CRISPR-Cpf1_Nterm** | 0.01 | 0.86 | 0.03 |
| **CRISPR-Cpf1_Cterm** | 0.09 | 0.73 | 0.16 |
| **CRISPR-Cpf1_Average** | 0.73 | 0.59 | 0.65 |

**Table 1 Out-of-class RNN classification performance** RNN models were trained toward the Ferritin, P450, Genome-Edit (CRISPR) functions using InterPro families/clusters of protein sequences. For Ferritin function, the bfr or dps family was left out as "test-set". Tripling the number of recurrent layers for a deeper model (Ferritin-dps_3X) increased recall for predicting dps. For P450, the B class or E class group II (E_II) was left out as "test-set". For CRISPR, the Cpf1 family was left out as "test-set". The average of the predictions on up to 800 amino acids in the N and C termini significantly increased precision and F1, suggesting several important features throughout the entire sequence that are necessary for function.